\documentclass[11pt,a4paper]{article}

\pdfoutput=1

\usepackage{jheppub}

\usepackage{amssymb,amsmath}
\usepackage{multicol,bbm}
\usepackage{pxfonts}
\usepackage{bm}
\usepackage{multirow}
\usepackage{xcolor}
\usepackage{mathrsfs}
\usepackage{graphicx}

\newcommand{\be}{\begin{equation}}
\newcommand{\ee}{\end{equation}}
\newcommand{\bea}{\begin{eqnarray}}
\newcommand{\eea}{\end{eqnarray}}

\newcommand{\TeV}{~\textrm{TeV}}
\newcommand{\GeV}{~\textrm{GeV}}

\title{Playing Tag with ANN:  Boosted Top Identification with Pattern Recognition}

\author[a]{Leandro G. Almeida,}
\author[b]{Mihailo Backovi\'{c},}
\author[c]{Mathieu Cliche,}
\author[d,e]{Seung J. Lee,}
\author[c]{Maxim Perelstein}

\emailAdd{almeida@biologie.ens.fr}
\emailAdd{mihailo.backovic@uclouvain.be}
\emailAdd{mc863@cornell.edu}
\emailAdd{sjjlee@kaist.ac.kr}
\emailAdd{mp325@cornell.edu}

\affiliation[a]{Institut de Biologie de l'\'Ecole Normale Sup\'erieure \,(IBENS),
Inserm 1024- CNRS 8197, \\ 46 rue d'Ulm, 75005 Paris, France}
\affiliation[b]{Center for Cosmology, Particle Physics and Phenomenology - CP3, Universite Catholique de Louvain, Louvain-la-neuve, Belgium}
\affiliation[c]{Laboratory for Elementary Particle Physics, Cornell University, Ithaca, NY 14853, USA}
\affiliation[d]{Department of Physics, Korea Advanced Institute of Science and Technology, 335 Gwahak-ro, Yuseong-gu, Daejeon 305-701, Korea}
\affiliation[e]{School of Physics, Korea Institute for Advanced Study, Seoul 130-722, Korea}

\abstract{
Many searches for physics beyond the Standard Model at the Large Hadron Collider (LHC) rely on top tagging algorithms, which discriminate between boosted hadronic top quarks and the much more common jets initiated by light quarks and gluons. We note that the hadronic calorimeter (HCAL) effectively takes a ``digital image" of each jet, with pixel intensities given by energy deposits in individual HCAL cells. Viewed in this way, top tagging becomes a canonical pattern recognition problem. With this motivation, we present a novel top tagging algorithm based on an Artificial Neural Network (ANN), one of the most popular approaches to pattern recognition. The ANN is trained on a large sample of boosted tops and light quark/gluon jets, and is then applied to  independent test samples. The ANN tagger demonstrated excellent performance in a Monte Carlo study: for example, for jets with $p_T$ in the $1100-1200$ GeV range, $60$\% top-tag efficiency can be achieved with a $4$\% mis-tag rate. We discuss the physical features of the jets identified by the ANN tagger as the most important for classification, as well as correlations between the ANN tagger and some of the familiar top-tagging observables and algorithms.     
}

\begin{document}

\preprint{{\scriptsize CP3-15-01\vspace*{.1cm}}}

\maketitle	
		
\section{Introduction} 
\label{sec:introduction}
Many extensions of the Standard Model (SM) predict new particles with masses around the TeV scale. Searches for such new particles form a major component of the experimental program at the Large Hadron Collider (LHC). In most models, the new particles are unstable, and their decays often contain weak-scale SM states, namely the $W$ and $Z$ bosons, the Higgs boson, and the top quark. Searches for final states containing top quarks are particularly important, due to the special role played by the top sector in many models of electroweak symmetry breaking. Decays of heavy new particles with mass above the electroweak scale typically result in highly energetic, relativistic top quarks in the lab frame. Identifying and characterizing such ``boosted" top quarks in the data is crucial for new physics searches and tests of naturalness~\cite{Perez:2013oaa} at the LHC, especially as the bounds on the new physics mass scales in many candidate models are pushed higher. Examples of new physics leading to boosted top signatures include Kaluza-Klein gluons~\cite{Agashe:2006hk,Lillie:2007yh} and string Regge states~\cite{Perelstein:2011ez} of the Randall-Sundrum model, stops~\cite{Plehn:2010st} and gluinos~\cite{Berger:2011af} of supersymmetry, top and light quark partner decays in Composite Higgs models~\cite{Azatov:2013hya,Flacke:2013fya, Backovic:2014uma, Backovic:2014ega, Gripaios:2014pqa,Reuter:2014iya}, and many others.


Due to relativistic kinematics, the decay products of a boosted top quark are highly collimated. For instance,  hadronic decay of a top quark of $p_T \sim 1 \TeV$ would produce three quarks collimated into a cone of rough size $R \sim 0.4$ and result in a specific pattern of hadronic activity in the detector. Classical event reconstruction techniques are inadequate to tag and measure such topologies, as most of the showered radiation falls into a small angular region. One solution is to cluster the event with a large jet cone ($R\sim 1$), and consider the features of energy distribution inside such ``fat" jets (so-called jet substructure), instead of correlations between individual small radius jets. Over the past decade, a variety of methods for boosted top tagging via jet substructure have been developed (see Ref.~\cite{Altheimer:2012mn} for a review), most of which can be cast into several (non exclusive) groups. Jet shapes are observables based on various moments of the jet energy distribution. Notable examples are angular correlations studied extensively in Ref.~\cite{Jankowiak:2011qa}, sphericity tensors~\cite{Bjorken:1969wi, Thaler:2008ju} and other perturbatively calculable jet shapes~\cite{Almeida:2008yp}. Considerations of jet clustering history led to development of numerous Filtering jet substructure methods~\cite{Butterworth:2008iy, Krohn:2009th,Ellis:2009me}, where the differences in the late steps of jet clustering between heavy SM states and QCD jets from light partons have been successfully applied in tagging of heavy SM states. Furthermore, Prong Taggers such as $N$-subjettiness~\cite{Thaler:2010tr,Thaler:2011gf} exploit the differences in the number of hard energy depositions within the boosted jet (e.g. three-body top decays compared to the typical two-body splitting of a light jet). Parton level models of boosted decays and kinematic constraints built into them can also be used to study jet substructure, with the Template Overlap Method (TOM)~\cite{Almeida:2011aa,Almeida:2010pa, Backovic:2012jj, Backovic:2013bga} being the most notable example. More recently, Matrix Element Method~\cite{Abazov:2004cs,Artoisenet:2010cn} inspired techniques such as Shower Deconstruction have emerged~\cite{Soper:2011cr,Soper:2012pb}, where a boosted jet is tagged using approximations to hard matrix elements and the parton shower. Soft drop declustering (a generalization of modified mass drop tagging) is another method which has been recently developed for removing non-global contributions (soft radiation) to the jet~\cite{Larkoski:2014wba}. Several of these methods have been implemented in the analyses of the LHC data by the CMS and ATLAS collaborations; see, for example Ref.~\cite{Aad:2012dpa, Aad:2012qa,CMS:2009lxa,CMS:2011xsa}.

In this paper, we pursue an alternative approach to jet substructure. Experimentally, information about hadronic activity in an event comes mainly from the hadronic calorimeter (HCAL), with the basic observable being the energy deposited in each of the HCAL cells. One can think of the information provided by the HCAL as a {\it digital image}, with each cell (or topo-cluster) being identified as a pixel, and with energy deposit in the cell corresponding to the intensity (or grayscale color) of that pixel. From this point of view, boosted top identification is simply a classic image-recognition problem: distinguishing the energy-deposit patterns characteristic of boosted tops from patterns due to other sources, such as the usual QCD jets. This suggests that computational algorithms developed in the field of image recognition could be of use in boosted top tagging.\footnote{Recently, Ref.~\cite{Cogan:2014oua} studied jet substructure as an image recognition problem in the context of boosted $W$ tagging as well gluon/quark discrimination. The authors utilised a linear Fisher discriminant  trained on a sample of signal and background events, in order to distinguish the desired events from the backgrounds. The method out-performs the existing methods of $W$ tagging, illustrating the benefits of the image recognition approach to jet substructure. }


With this motivation, we constructed a new top tagger algorithm based on one of the most popular approaches to image recognition, Artificial Neural Networks (ANNs). In this approach, each jet is classified as top or non-top according to a highly non-linear scoring function. The function contains multiple adjustable parameters, called weights. These are chosen using a training procedure, in which the ANN is presented with a large sample of jets that are known to be top or non-top, and the weights are chosen to maximize the number of correctly identified jets in this sample. (In our study, all samples are generated by Monte Carlo simulations. In experimental applications, ANN may be trained on either MC samples or carefully selected ``calibration" data sets.) Having fixed the weights, the ANN is then applied to independent samples containing both top and non-top jets, and asked to discriminate between them. We find that the performance of the ANN tagger significantly exceeds that of several popular tagging algorithms currently in use over a wide range of $p_T$, demonstrating the practical utility of this approach. 

The paper is organized as follows. Section~\ref{sec:evtgen} describes the MC event samples used for training and testing the ANN tagger, as well as the pre-processing steps applied to these samples before the ANN is applied. Section~\ref{sec:neuraljet} contains a detailed description of the ANN tagger, including the network architecture and the training algorithms we employed. In Section~\ref{sec:results}, we present the results of our study of ANN tagger performance and comparisons with other popular taggers. We also discuss the physical features of jets that are dominant in the ANN classification, and the extent to which ANN output is correlated with that of other taggers. We conclude with a recap and a brief discussion of directions for future research in Section~\ref{sec:discussion}. An Appendix contains a brief description of the top taggers we use for the purpose of comparison with the ANN tagger.

\section{Event Generation and Pre-Processing}
\label{sec:evtgen}
We generate benchmark event samples with \verb|MadGraph 5|~\cite{Maltoni:2002qb} at leading order, and shower them with  \verb|Pythia 6|~\cite{Sjostrand:2006za}. In order to study the effects of different showering algorithms on the results, we also generate separate data samples showered with \verb|Pythia 8|~\cite{Sjostrand:2007gs}. For simplicity, we extract a pure sample of top jets from a Standard Model top pair-production simulation, at leading order with no matching. The tops are decayed in \verb|MadGraph 5|, so that the angular distribution of the decay products is modeled correctly.
Similarly, we generate the light jet sample from a simulation of the QCD di-jet process, including both quarks and gluons in the final state, but no matching to extra jets. Fiducial cut $|\eta|\leq 5.0$ is imposed at the hadron level. We cluster the events using the \verb|fastjet|~\cite{Cacciari:2011ma} implementation of the anti-$k_T$ algorithm~\cite{Cacciari:2008gp} with a large jet cone of $R=1.0$. For our analysis, we only use the highest $p_T$ jet in each event, and impose the cut $|\eta_{\rm jet}|\leq 2.5$. We consider samples of jets within three jet $p_T$ ranges: $500-600$ GeV, $800-900$ GeV and $1100-1200$ GeV. These three bins span a range of jet $p_T$ values relevant for top tagging at the LHC, while analyzing them separately provides information about $p_T$ sensitivity of the tagging efficiency and other parameters. Unless otherwise noted, we impose a cut on the jet mass ({\it i.e.} the invariant mass of all particles assigned to the jet), selecting jets within a window
\be
	130 \GeV < m_J^{R=1.0} < 210 \GeV.
	\label{eq:mjet}
\ee 
A vast majority of top jets fall within this mass range, while most QCD jets are rejected by this cut. Discriminating the remaining QCD jets from top jets is the task for the top tagger.

In order to form an input to the ANN tagger, we preprocess each jet as follows. First, we find the center of the jet, defined by the sum of the coordinates of all particles weighted by their energies, 
\be
\eta_C = \frac{1}{E}\sum_j \eta_j E_j,~~\phi_C = \frac{1}{E}\sum_j \phi_j E_j,
\ee 
where $E=\sum_j E_j$ is the total energy of the jet. We then shift the coordinates of each particle so that the jet is centered at the origin in the new coordinates: 
\be
\eta_j^\prime = \eta_j - \eta_C,~~\phi_j^\prime = \phi^j - \phi_C.
\ee
Further, we find the jet ``principal axis" in the $(\eta,\phi)$ plane, defined by
\begin{eqnarray}
\tan(\theta) &=& \frac{\sum_{j}\frac{\phi_j^\prime\cdot E_j}{\Delta R^\prime}}{\sum_{j}\frac{\eta^\prime_j\cdot E_j}{\Delta R^\prime}},~~~~~\Delta R^\prime=\sqrt{\eta_j^{\prime2}+\phi_j^{\prime2}},
\end{eqnarray}
and rotate the coordinate system so that this principal axis is the same direction ($+\eta$) for all jets:
\begin{eqnarray}
\eta_j^{\prime\prime} &=& \eta_j^\prime\cdot\cos(\theta)+\phi_j^\prime\cdot\sin(\theta),\\
\phi_j^{\prime\prime} &=& -\eta_j^\prime\cdot\sin(\theta)+\phi_j^\prime\cdot\cos(\theta).
\label{eq:rot}
\end{eqnarray}
These coordinate transformations remove information about the jet position in the calorimeter and its orientation in the $(\eta, \phi)$ plane. Both pieces of information are irrelevant for top tagging, and removing them from consideration allows the ANN tagger to focus on the irreducible physical differences between top and QCD jets.\footnote{As an exercise, we also attempted to train the neural network on a set of jets with randomly oriented principal axes, {\it i.e.} without the rotation~(\ref{eq:rot}). We found that this procedure still yields an effective tagger; presumably, the neural net learns to ignore the axis orientation information during the training process. However, to achieve the same tagging performance, the randomly-oriented training set needs to be significantly larger.}  


In the new coordinates, nearly all (98\%) of the particles assigned to a given jet fall within a window of $\eta^{\prime\prime}\in [-\pi/2, \pi/2]$ and $\phi^{\prime\prime}\in [\pi/2, \pi/2]$. We model the HCAL response to the jet by dividing this window into $30\times 30$ square cells. (The cell size is approximately $0.1\times 0.1$, close to the realistic values in ATLAS and CMS.) The normalized energy deposited in each cell, $\varepsilon_{ab}$ ($a, b=1\ldots 30)$, is computed by adding up the energies of all particles falling within that cell, and dividing by the total energy of the jet. (The last step is once again necessary to remove information irrelevant for top tagging, in this case the total jet energy.) By construction, $\varepsilon_{ab}$ is dimensionless and lies between 0 and 1. In the language of image processing, each jet has been converted into an image with $30\times 30$ pixels, with a grayscale color of each pixel given by the corresponding $\varepsilon_{ab}$. These images can now be classified by an Artificial Neural Network (ANN), described in the following section.

\section{ANN Tagger}
\label{sec:neuraljet}
\begin{figure}[t]
\begin{center}
\includegraphics[width=1.0\textwidth]{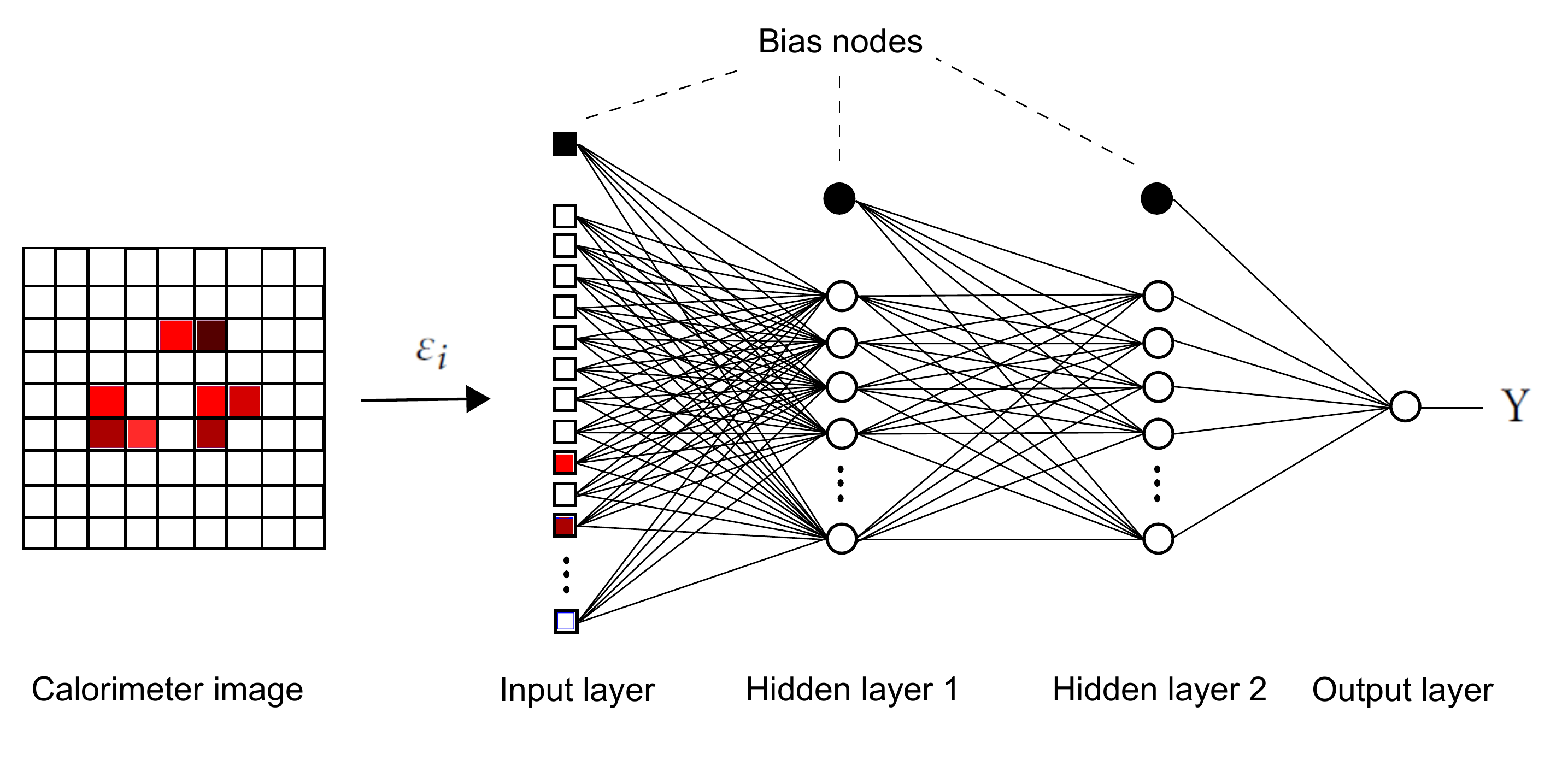}
\end{center}
\caption{Graphical representation of the Artificial Neural Network (ANN).
\label{fig:neural_net}}
\end{figure}
 
ANN tagger is based on a feed-forward neural network with an input layer consisting of $30\times 30=900$ nodes, one for each calorimeter cell; two hidden layers, of 100 nodes each, to process the signal; and an output layer consisting of a single node, whose value $Y$ is interpreted as the probability that a given jet comes from a boosted top decay. The architecture of the network is shown in Fig.~\ref{fig:neural_net}. (For pedagogical introduction to Artificial Neural Networks in the context of image recognition, see for example~\cite{Bishop}.) Mathematically, the ANN can be thought of as a succession of non-linear transformations:\footnote{In Eq.~(\ref{NNformula}) and below, repeated indices are always summed over.}
\begin{eqnarray}
\epsilon_i \to h^{(1)}_i = f( W^{(1)}_{ij} \epsilon_j +b^{(1)}_i) \to \cdots \to  h^{(l)}_i = f( W^{(l)}_{ij} h^{(l-1)}_j +b^{(l)}_i ) \to  Y = f( W^{(O)}_{j} h^{(l)}_j +b^{(O)}),
\label{NNformula}
\end{eqnarray}
where $f$ is the so-called activation function, chosen to be 
\begin{equation}
f(z) = \frac{1}{1+e^{-z}}.
\end{equation}
The inputs $\epsilon_i$ are simply the normalized energy deposits $\varepsilon_{ab}$ defined above, rearranged in a single 900-dimensional vector: $\varepsilon_{ab}\equiv\epsilon_{30a+b}$. The weights $W^{(L)}_{ij}$ and the biases $b^{(L)}_i$ are numbers determined by the training procedure, which we will now describe.
 
To train the network, we use a set of $N/2$ top and $N/2$ QCD jets, where $N$ is a large number. For the $i$-th jet, we assign the ``target output" variable: $y_i=1$ if it is a top jet, and $y_i=0$ if it is a QCD jet. Training consists of adjusting the weights so that the actual outputs of the ANN $Y_i$ correspond as close as possible to the target outputs $y_i$, across the training set. To quantify the error, we use 
the logarithmic loss variable
\begin{eqnarray}
\text{Log-loss} = -\frac{1}{N}\sum_{i=1}^{N}\left[y_i\log(Y_i)+(1-y_i)\log(1-Y_i)\right].
\end{eqnarray}
The goal of training is to choose weights that minimize this function. We use the back-propagation algorithm~\cite{Werbos:1975}, combined with gradient-descent minimization. In its simplest version, the algorithm can be summarized as follows~\cite{Marsland:2009:MLA:1571643}:
\begin{enumerate}
\item Initialize the weights of each link to small random values.  
\item Repeat until convergence of log-loss, for each input vector $\epsilonup_i$: 
\begin{itemize}
\item Forward: Compute the output of each neuron until the output layer is reached, that is 
\begin{eqnarray}
\epsilon_i \to h^{(1)}_i = f( W^{(1)}_{ij} \epsilon_j +b^{(1)}) \to   h^{(2)}_i = f( W^{(2)}_{ij} h^{(1)}_j +b^{(2)}) \to  Y = f( W^{(O)}_{j} h^{(2)}_j )
\end{eqnarray}
\item Backward: Adjust the weights of each neuron by propagating backward the error at the output using
\begin{eqnarray}
\delta^{(O)} &=& (y-Y)Y(1-Y) \text{  and  } \delta^{(l)}_{k}=h^{(l)}_k(1-h^{(l)}_k)\sum_{j} W^{(l-1)}_{kj} \delta^{(l-1)}_{j}\\
W^{(0)}_{k} &\rightarrow& W^{(0)}_{k} +\eta \delta^{(O)} h^{(2)}_k \nonumber\\
W^{(2)}_{jk} &\rightarrow& W^{(2)}_{jk} +\eta \delta^{(2)}_j h^{(1)}_k \nonumber\\
W^{(1)}_{jk} &\rightarrow& W^{(1)}_{jk} +\eta \delta^{(1)}_j \epsilonup_k 
\end{eqnarray}
where $\eta$ is a small parameter called the learning rate.
\end{itemize}
\end{enumerate}   
We used several well-known tricks to make this algorithm more efficient.  First, instead of updating the weights after each jet $\epsilonup_i$, we used what is known as batch gradient descent so that the update on the weights is only done after all the jets in the training set have been processed.  In that scenario, the updates on the weights are an average of the individual updates caused by each jet.  Moreover, to reduce the odds of getting stuck at local minima we add what is known as a ``momentum" to the updates.  This means that the weights at iteration $t$, $W^{t}_{ij}$, are still being pushed by the update from the previous iteration $\Delta W^{t-1}_{ij}$, for example
\begin{eqnarray}
W^{t}_{ij} &\rightarrow& W^{t}_{ij} + \eta \delta^{(l)}_i h^{(l-1)}_j + \alpha \Delta W^{t-1}_{ij}
\end{eqnarray} 
where $\alpha\in (0,1)$ is a fixed parameter.  

\begin{figure}[t]
\begin{center}
\includegraphics[width=0.50\textwidth]{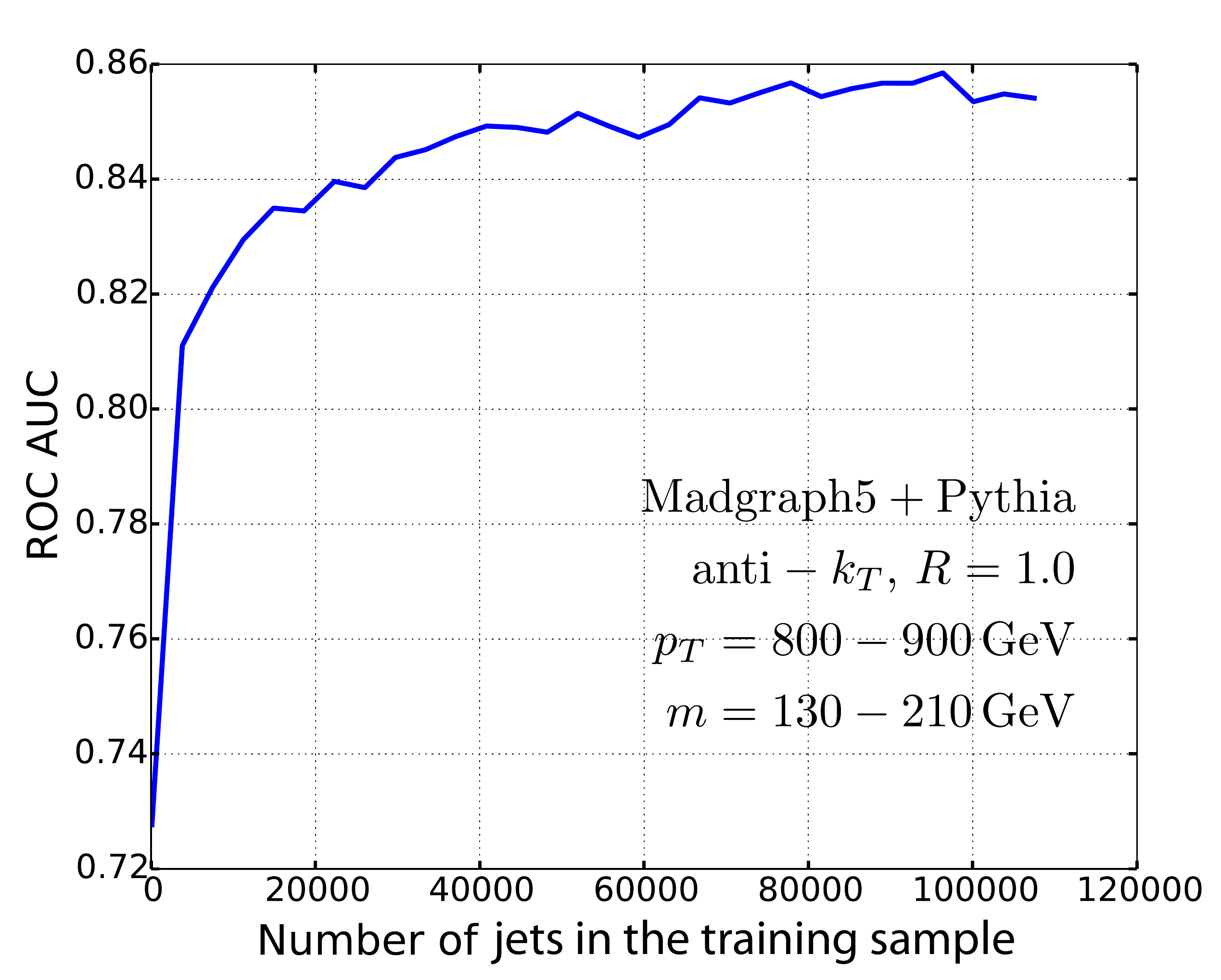}
\end{center}
\caption{ROC AUC on a cross-validation set of 50 000 jets, vs. number of jets in the training set.
\label{fig:train_size}}
\end{figure}

A major concern in using ANN classifiers is over-fitting the network to the training data. Over-fitting is a common problem in machine learning, in which the training procedure produces a classifier that emphasizes random fluctuations in the training data set, as opposed to the underlying trend. An over-fitted classifier would achieve excellent performance on the training set, but this will not generalize well to data sets which were not part of the training set, rendering it useless. Many techniques for avoiding over-fitting have been proposed in the literature. However, over several experiments we found that it was easier to avoid over-fitting simply by using more training data and ensembling several neural networks together.  To determine the size of the training set $N_{\rm tr}$ needed to saturate the learning of our neural network, we studied the performance of the trained network on a cross-validation set of 50000 top and QCD jets, as a function of $N_{\rm tr}$. For this analysis, the performance is characterized by the ROC AUC (area under the receiver operating characteristic curve) performance metric, which assigns a value of 0.5 to a random classifier and a value of 1.0 to a perfect classifier. As can be seen on Fig.~\ref{fig:train_size}, performance steadily improves with the training set size until $N_{\rm tr}\approx40 000$ ({\it i.e.} 20000 top images and 20000 dijet images), after which convergence is achieved. This indicates minimal over-fitting beyond that point. 

To further improve the performance of our tagger, we ensembled multiple neural networks together. The idea is to train $B$ neural networks together, with the output given by the average of their outputs,
\be
{\cal O}= \frac{1}{B}\sum_{i=1}^BY_i.
 \ee 
 In our application, $B=10$. All networks are trained using the same training set, but the jets are weighted. For the first network, all weights are set to one. Jets which are heavily misclassified by the first network are then assigned a larger weight, while jets which are correctly classified are assigned a smaller weight.  This re-weighted training set is then used to train the second network, and so on. This procedure allows the training algorithm to focus on specific events that are particularly arduous to classify, improving overall performance. For some parameter choices, this method can be mapped to boosted methods such as ADAboost~\cite{Freund99ashort}, where the weak classifiers are feed-forward ANNs.

\section{Results}
\label{sec:results}
\begin{figure}[t]
\begin{center}
\includegraphics[width=0.57\textwidth]{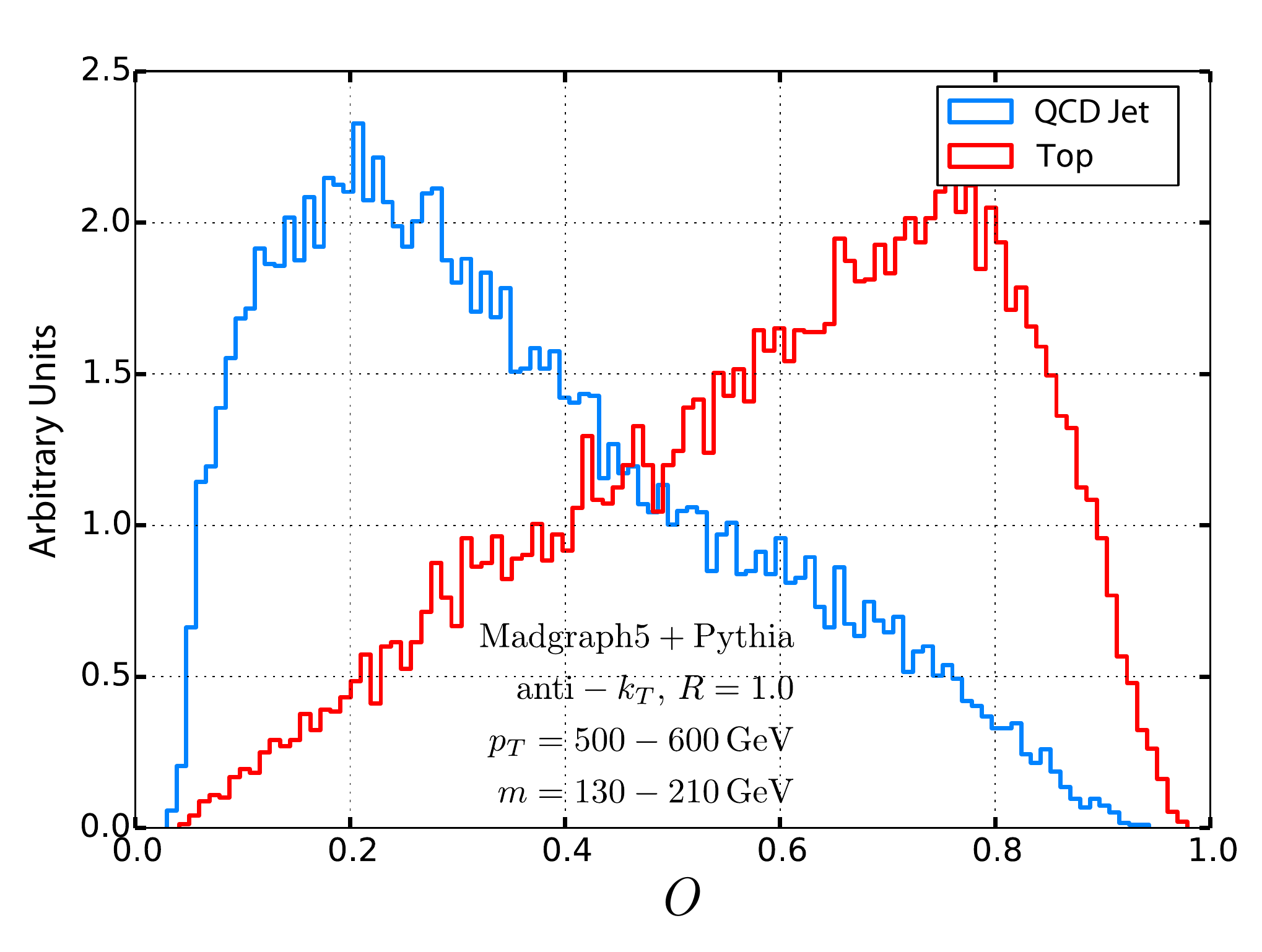}
\includegraphics[width=0.57\textwidth]{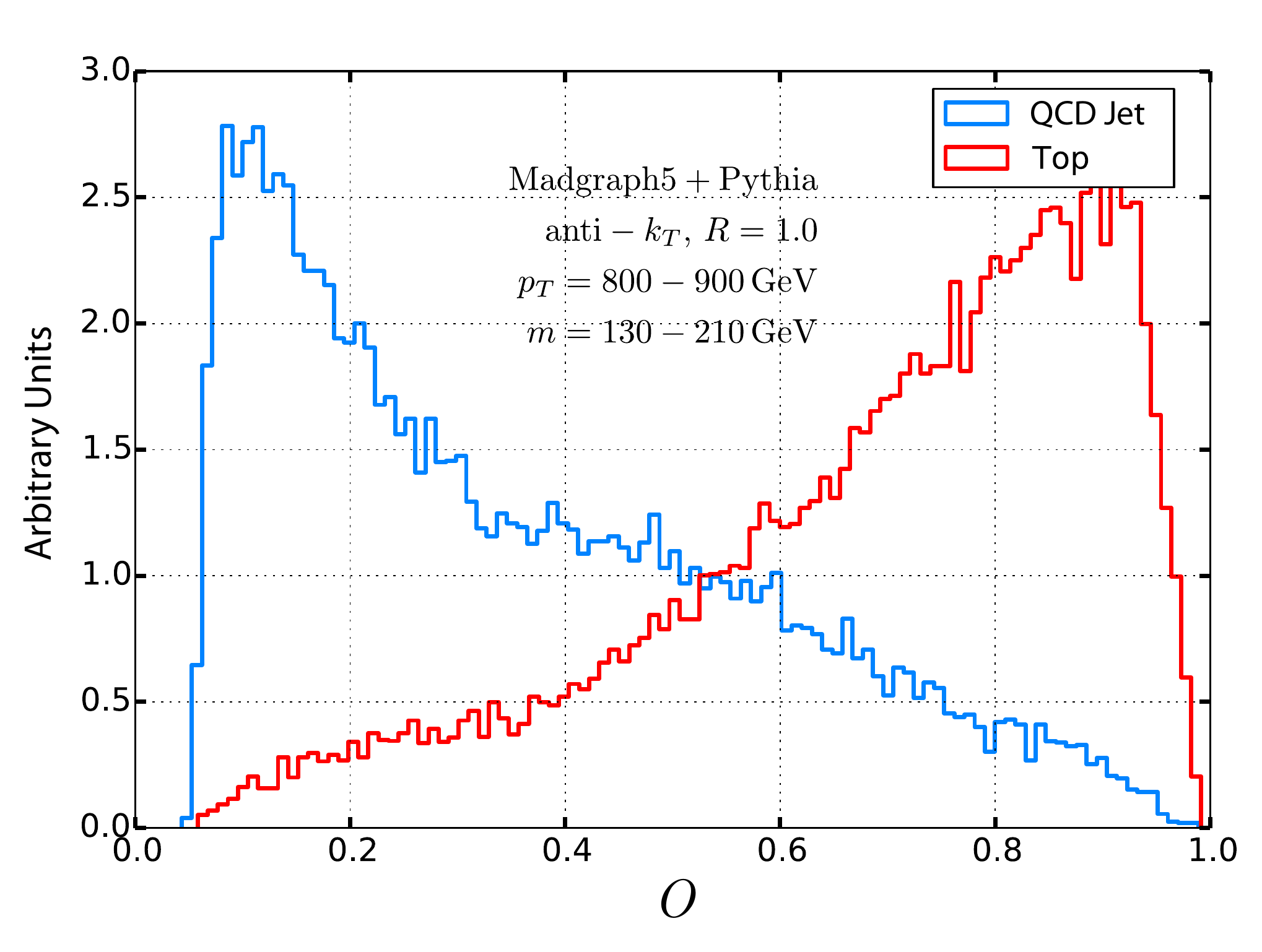}
\includegraphics[width=0.57\textwidth]{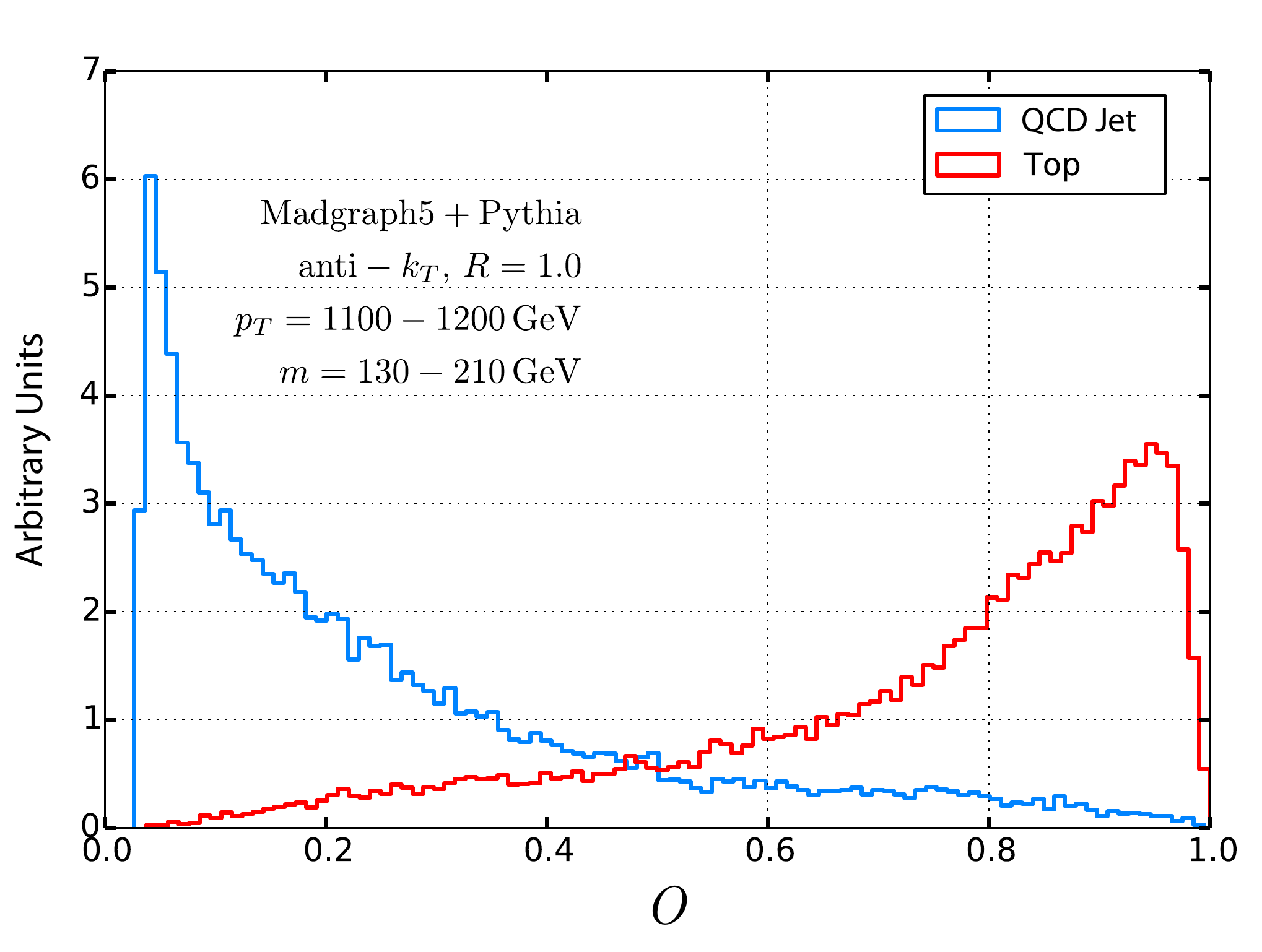}
\end{center}
\caption{Distributions of the ANN output ${\cal O}$ on top (red) and QCD (blue) jet samples in three representative $p_T$ ranges. All distributions are normalized to unit area. 
\label{fig:Odist}}
\end{figure}

The ensemble of ANNs described above has been trained on sets of about 50,000 top and QCD jets each, in three $p_T$ bins, $500-600$ GeV, $800-900$ GeV, and $1100-1200$ GeV. It has then been applied to test sets consisting of about 15,000 top and QCD jets each, in the same $p_T$ bins. The distribution of the neural network output ${\cal O}$ on the test sets is shown in Fig.~\ref{fig:Odist}. The classification power of this observable is clear from the figure: top jets are predominantly assigned ${\cal O}\approx 1.0$, while QCD jets are predominantly assigned ${\cal O}\approx 0.0$. To use the ANN ensemble as a top-tagger, we simply choose a threshold value ${\cal O}_{\rm th}$, and assign the ``top tag" to any jet with ${\cal O}\geq{\cal O}_{\rm th}$ and the ``QCD tag" to any jet with ${\cal O}<{\cal O}_{\rm th}$.

\begin{figure}[h!]
\begin{center}
\includegraphics[width=0.52\textwidth]{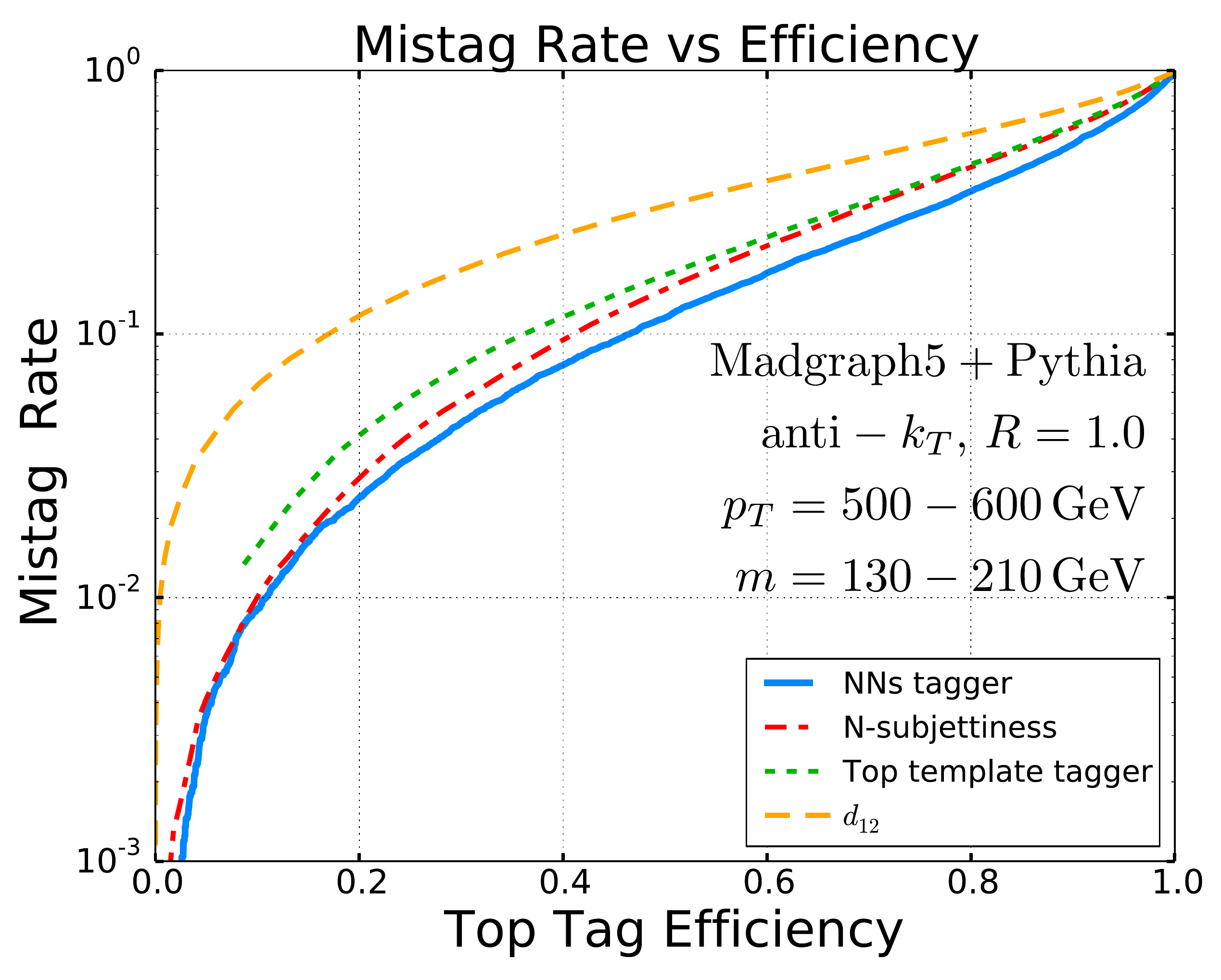}
\includegraphics[width=0.52\textwidth]{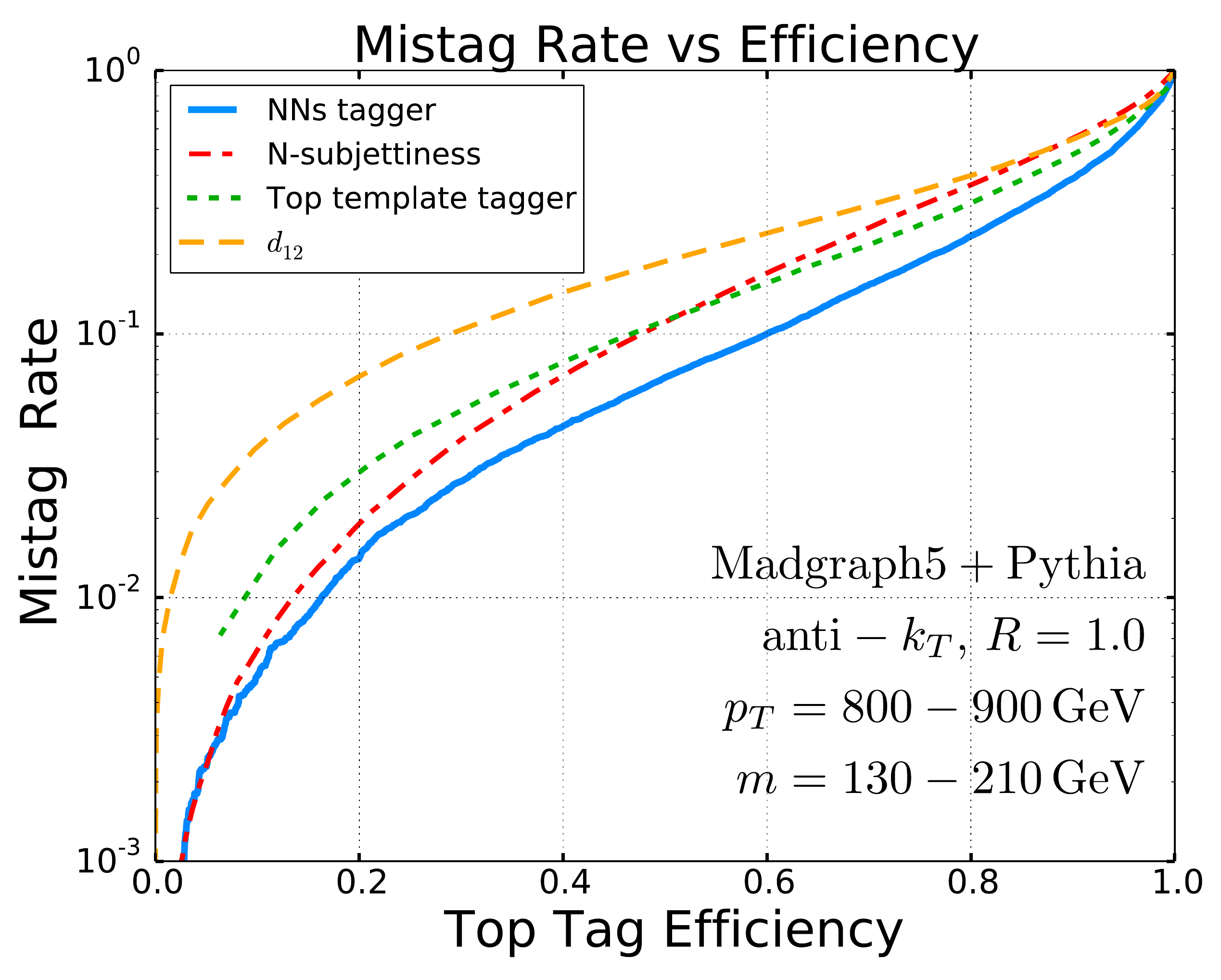}
\includegraphics[width=0.52\textwidth]{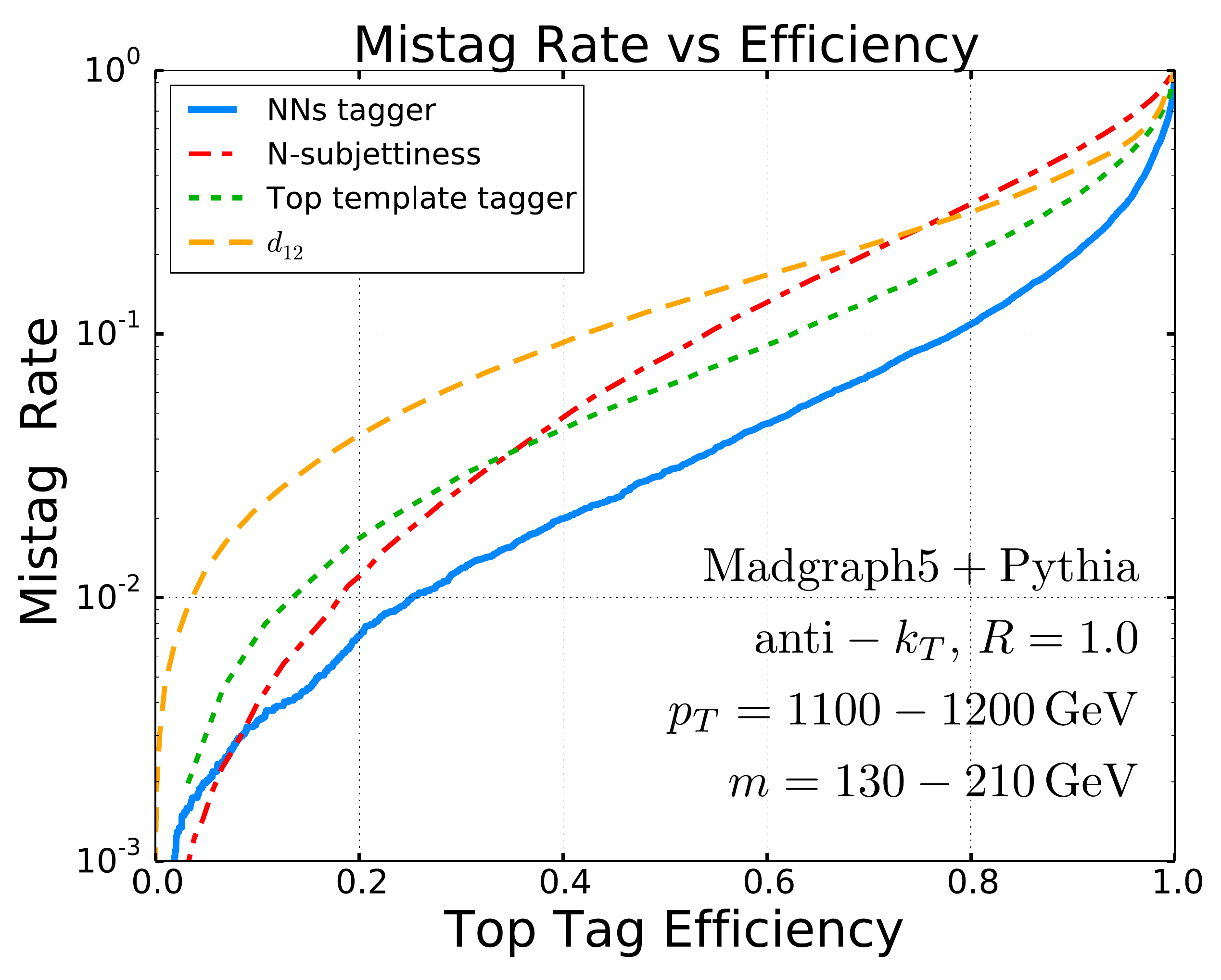}
\end{center}
\caption{Efficiency vs. Mis-tag rate curves for the ANN tagger (blue/solid lines), for jets in three representative $p_T$ ranges. For comparison, corresponding curves for three existing top taggers are also shown: 
$d_{12}$ tagger (yellow/dashed), top template tagger (green/dotted), and N-subjettiness (red/dash-dotted). 
\label{fig:EFFvsMT}}
\end{figure}

To discuss the performance of the ANN tagger, it is convenient to define efficiency and mis-tag rates as follows:
\be
{\rm Eff} = \frac{N^{\rm top}_{\rm top}}{{\cal N}_{\rm top}},~~~{\rm Mistag} = \frac{N^{\rm top}_{\rm QCD}}{{\cal N}_{\rm QCD}},
\label{EffMTdef}
\ee
where ${\cal N}_{\rm top}$ and ${\cal N}_{\rm QCD}$ are the total number of jets in the top and QCD jet samples, respectively, and $N^b_a$ is the number of jets in sample $a$ tagged as jets of type $b$ ($a, b=$top, QCD). Efficiency and mis-tag rates can be varied by varying the threshold ${\cal O}_{\rm th}$. The performance of the ANN tagger is shown in Fig.~\ref{fig:EFFvsMT}, where for comparison we also show the performance of three representative existing taggers, described in the Appendix. In all cases, the ANN tagger outperforms the existing taggers, achieving lower mis-tag rates for the same tagging efficiency. The improvement is especially dramatic for high jet $p_T$: for example, for jets with $p_T\in[1.1, 1.2]$ TeV range, the ANN tagger achieves 60\% tagging efficiency with about 4\% mis-tag rate, about a factor of 2 lower than the best of the existing taggers in our comparison pool. This clearly demonstrates the promise of the ANN-based approach.  

\begin{figure}[t]
\begin{center}
\includegraphics[width=1.0\textwidth]{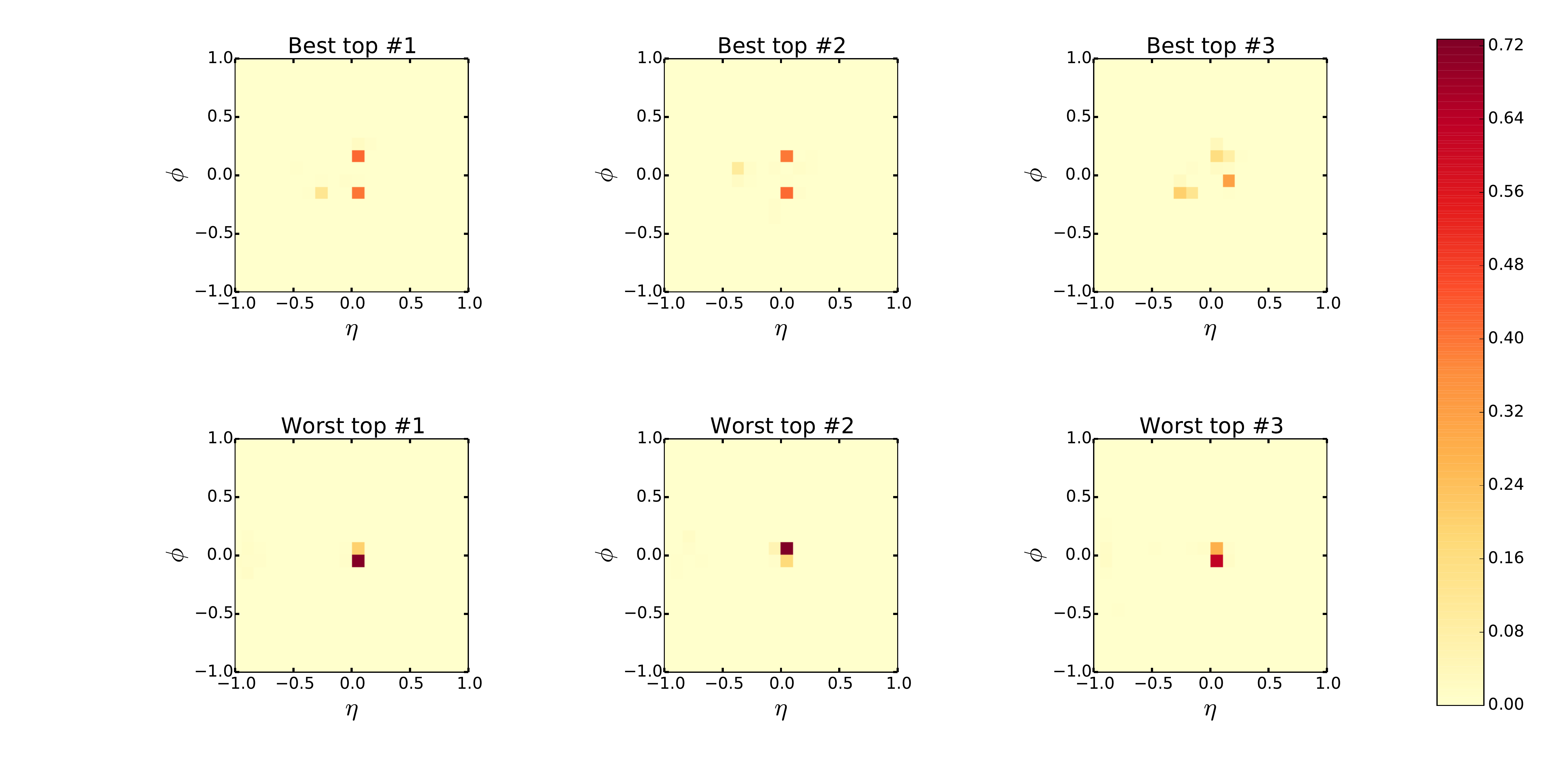}
\end{center}
\caption{Energy deposit patterns for three jets with the highest (top row) and lowest (bottom row) ANN scores in the top sample with $p_T\in[800, 900]$~GeV. 
\label{fig:top_images}}
\end{figure}

\begin{figure}[t]
\begin{center}
\includegraphics[width=1.0\textwidth]{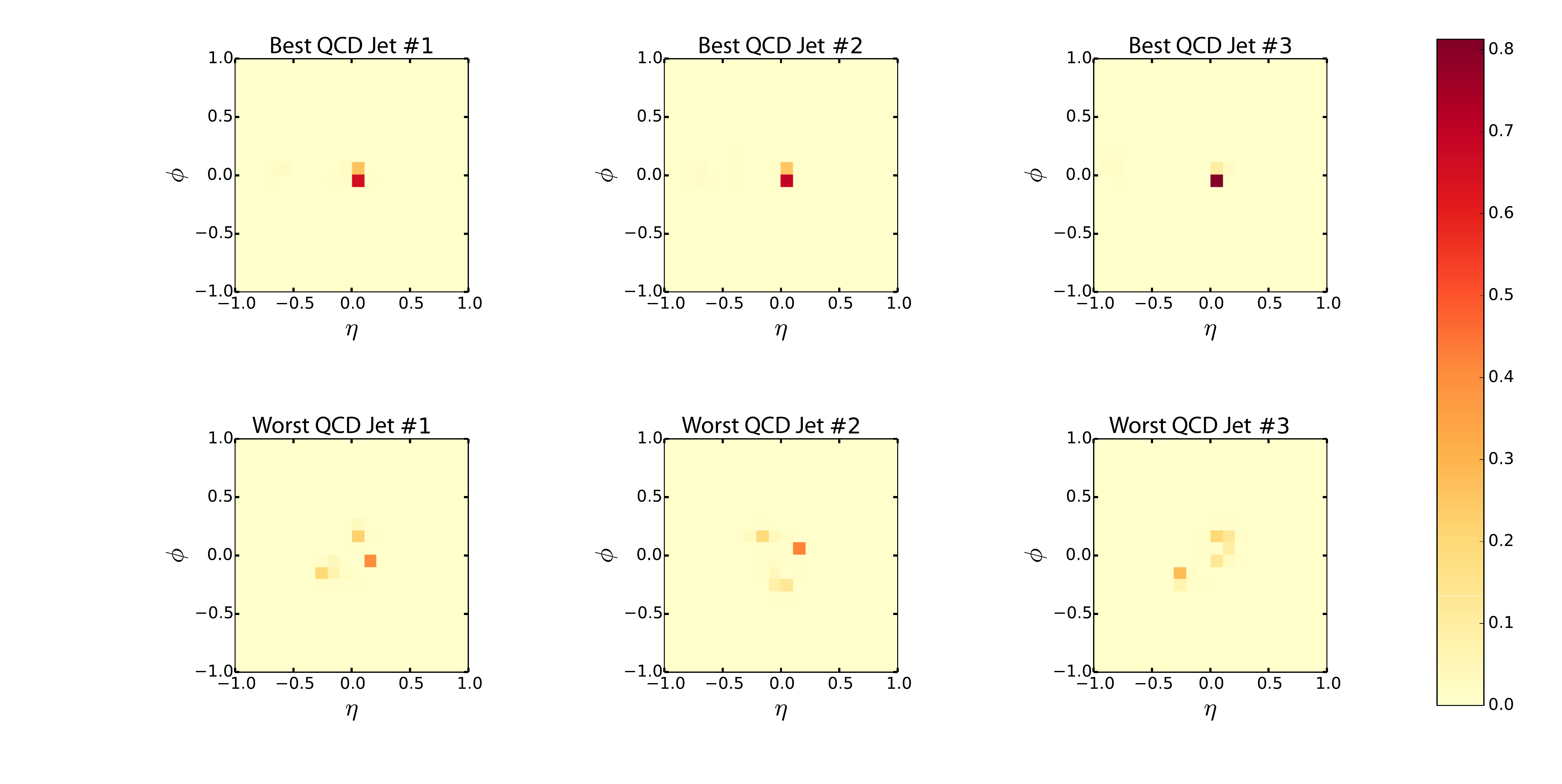}
\end{center}
\caption{Energy deposit patterns for three jets with the lowest (top row) and highest (bottom row) ANN scores in the QCD jet sample with $p_T\in[800, 900]$~GeV. 
\label{fig:QCD_images}}
\end{figure}

What physical features of the jet are identified by the ANN as the primary characteristics of a top jet? Some insight is provided by the energy deposit patterns of the highest-scoring and lowest-scoring jets, according to the ANN output ${\cal O}$, in the top sample. These are shown in Fig.~\ref{fig:top_images}. It is clear that the jets receiving high scores are characterized by well-defined three-prong structure, with each of the three quarks from top decay forming a well-defined, relatively isolated subjet.  The lowest-scoring jets are those where either the quarks are nearly collinear, or one of them is much softer than the other two (in the detector frame). Likewise, the QCD jets receiving the highest scores, and thus most likely to be mis-identified as tops, have well-defined, isolated subjets, while the QCD jets correctly tagged as such do not: see Fig.~\ref{fig:QCD_images}. 

\begin{figure}[t]
\begin{center}
\includegraphics[width=0.49\textwidth]{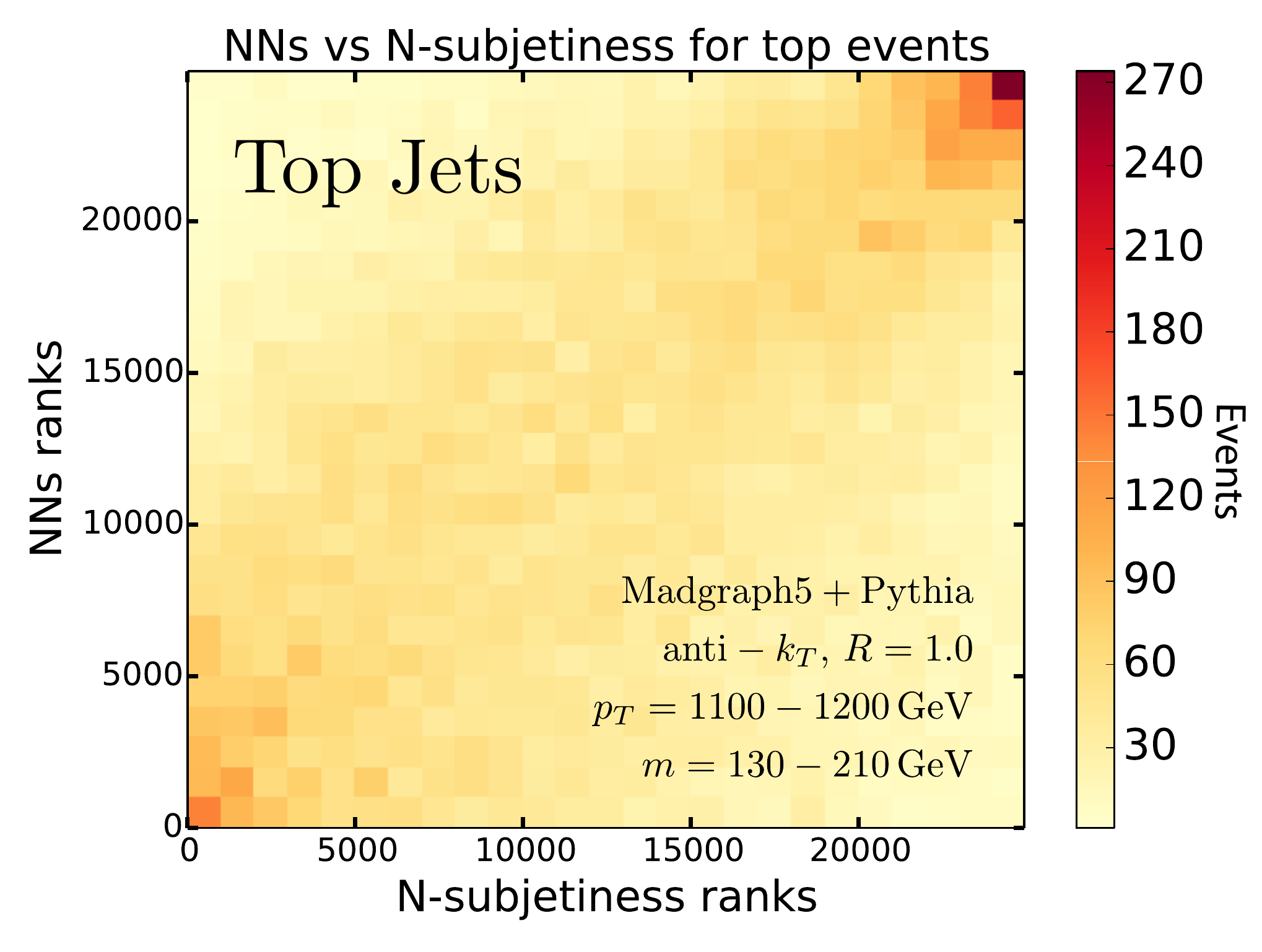}
\includegraphics[width=0.49\textwidth]{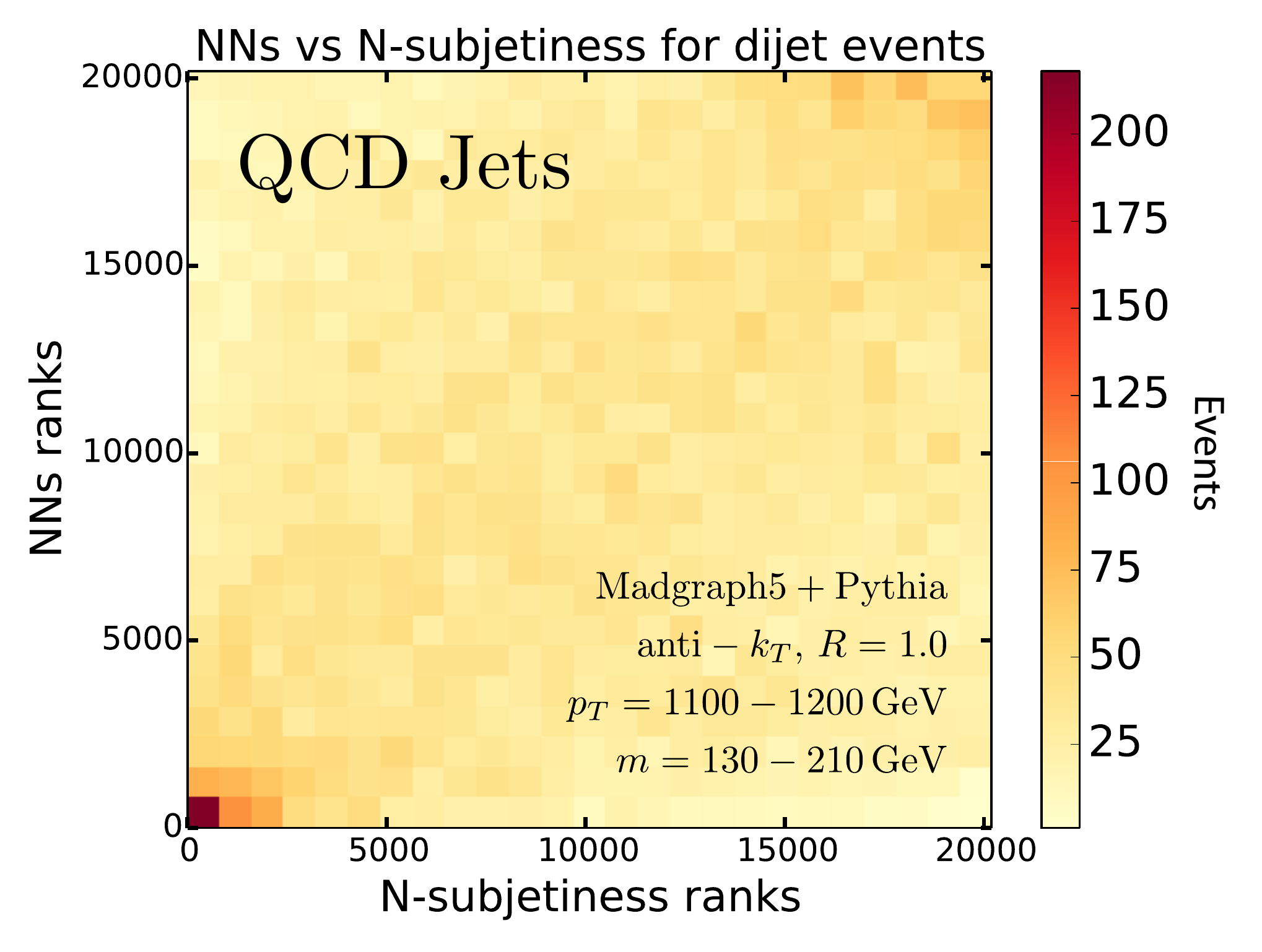}
\end{center}
\caption{Correlation between the rankings of jets according to $N$-subjettiness (horizontal axis) and ANN score (vertical axis). Left: top sample, $p_T\in [1100, 1200]$ GeV. Right: QCD jet sample, same $p_T$ range. Jets are ranked in order of increasing ``topness" for both samples. 
\label{fig:cor}}
\end{figure}

\begin{table}[h]
\begin{center}
\begin{tabular}{|l|r|r|r|r|}
\hline
Tagger & \multicolumn{2}{|c|}{Top} & \multicolumn{2}{|c|}{Dijet} \\
\cline{2-5}
 & $p_T\in[500, 600]$ & $p_T\in[1100, 1200]$ & $p_T\in[500, 600]$ & $p_T\in[1100, 1200]$ \\
\hline
TOM & 0.50 & 0.52 & 0.52 & 0.65 \\
$N$-sub. & 0.59 & 0.52 & 0.48 & 0.31 \\
ATLAS & 0.33 & 0.44 & 0.42 & 0.72 \\
\hline
\end{tabular}
\end{center}
\caption{Correlation coefficients between the ANN score and the output of alternative taggers, in a variety of samples.}
\label{tab:cor}
\end{table}%

To gain further insight, we studied correlations of the ANN scores with other observables used to tag tops. Table~\ref{tab:cor} contains the correlation coefficients between the ANN score and the output of the other taggers in our comparison pool, on a variety of samples used in our analysis. (The correlation coefficients are normalized so that $1.0$ indicates perfect correlation and $-1.0$ perfect anti-correlation, while $0$ indicates absence of correlation.) In all cases, we observe significant, though far from perfect, positive correlations, with coefficients ranging from about 0.3 to 0.7. A visual illustration is provided by Fig.~\ref{fig:cor}, which shows that the ranking of jets according to the ANN score and the $N$-subjettiness are indeed correlated, in both top and light-jet samples; correlation plots for all other taggers and $p_T$ ranges look very similar. This should not be surprising since all top taggers to some extent exploit the same physical characteristics of the boosted top jets. Nevertheless, as noted above, ANN systematically outperforms the other taggers in terms of tagging efficiency vs. mistag rates, indicating that the complicated non-linear observable created by the ANN learning process captures the information present in the jet substructure in a more optimal way. In other words, it seems that all taggers find roughly the same subset of jets to be ``easily classifiable", and all have a very good success rate on this subset. However, the ANN tagger seems to be able to correctly classify a higher fraction of the jets outside of this subset, leading to higher overall success rate.

\begin{figure}[t]
\begin{center}
\includegraphics[width=0.49\textwidth]{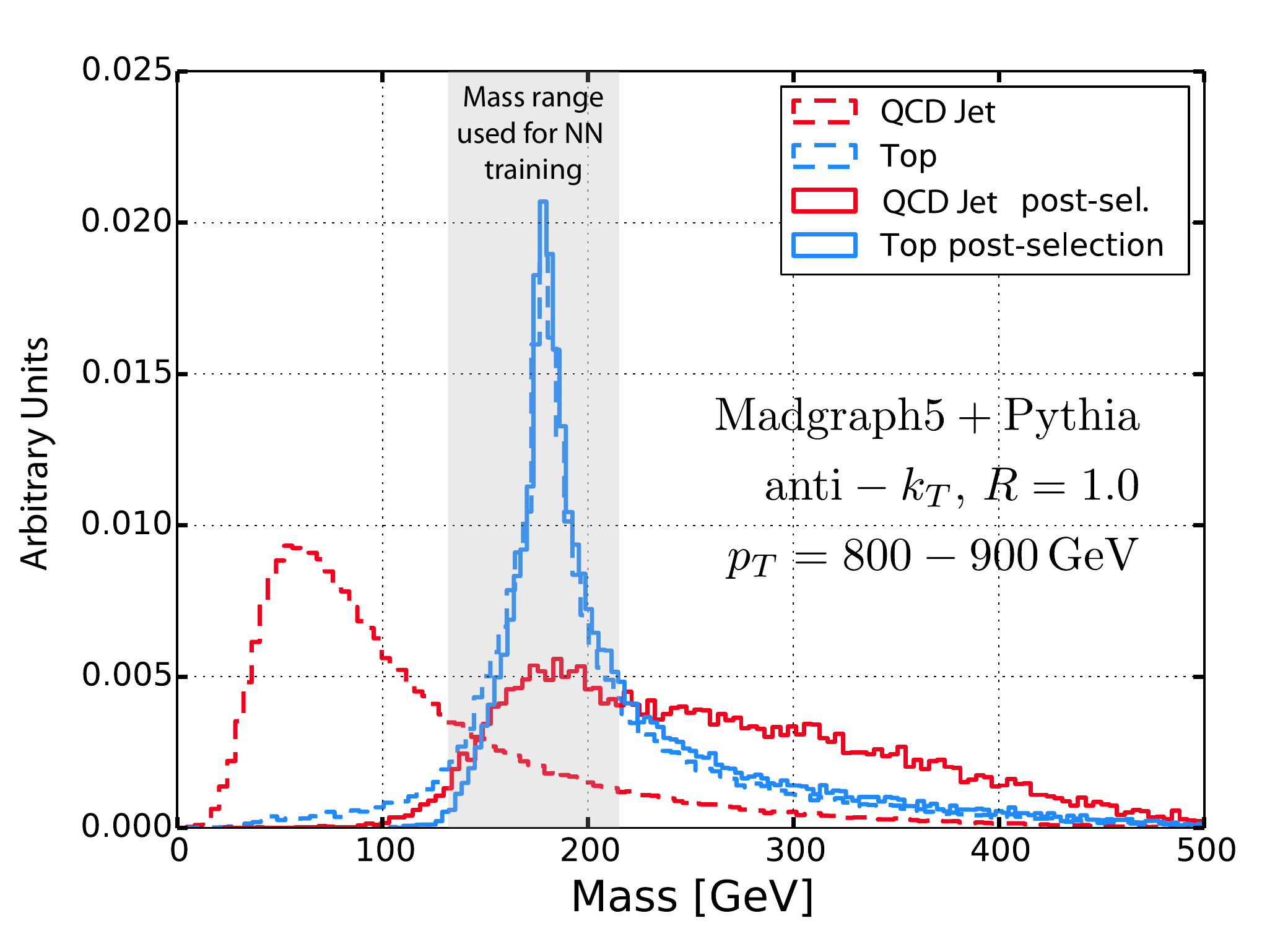}
\includegraphics[width=0.49\textwidth]{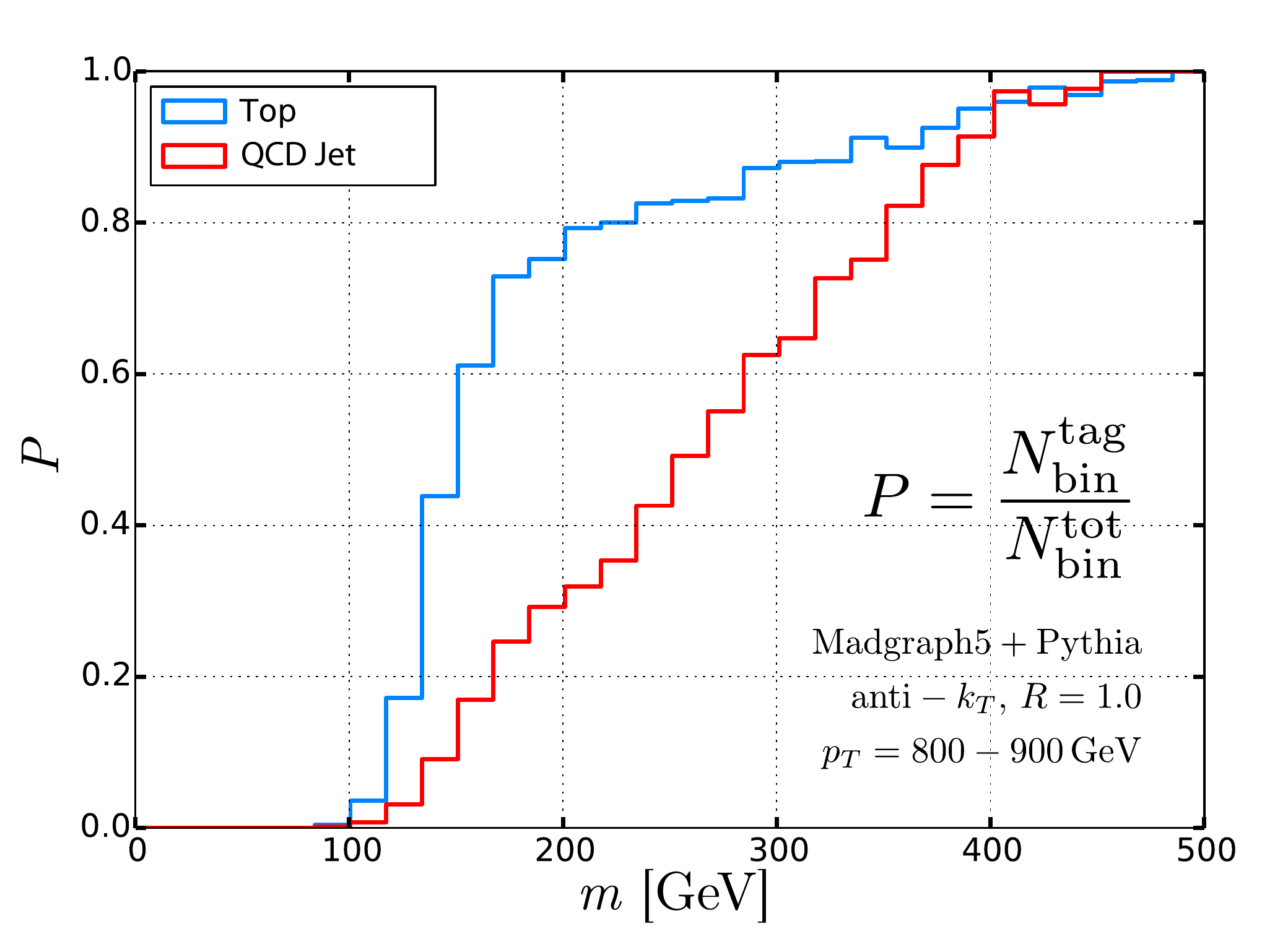}
\end{center}
\caption{Left: Jet mass distributions for top (blue) and dijet (red) samples with $p_T\in[800, 900]$ GeV window, and no mass cut. Dashed lines: all jets; solid lines: jets tagged as tops by the ANN tagger. All distributions are normalized to unit total area. Right: probabilities for a jet in the top (blue) and dijet (red) samples to be tagged as a top jet by the ANN tagger.
\label{fig:mass}}
\end{figure}

Another interesting question is how the ANN performance varies with the jet mass. The training samples and test samples in all plots shown so far only contain jets in a $130\ldots 210$ GeV mass window, where most top jets are expected to lie. We also applied the ANN tagger to the full sample of jets in the $[800, 900]$ GeV $p_T$ range, without the mass cut. The jet mass distributions in this sample, before and after the ANN tagger is applied, as well as the tagging probability as a function of the jet mass, are shown in Fig.~\ref{fig:mass}. (The cut on the ANN output used in the figure corresponds to the overall tag efficiency in the $130\ldots 210$ GeV mass window of 70\%.) For jet mass below 130 GeV, the probability of a positive top tag drops rapidly, for both top and QCD jets.  This is presumably due to the fact that jets with a clear three-prong structure are unlikely to have a low mass.  On the other hand, for jet mass above 210 GeV, the probability of a positive top tag is roughly independent of the jet mass. It should also be noted that the tag probability is smooth on the boundaries of the mass window selected for training, indicating that there is no strong dependence on the choice of the training sample. The ability of the ANN tagger to reject jets with low invariant mass may be useful in reducing effects of the pile-up. 

\begin{figure}[t]
\begin{center}
\includegraphics[width=0.49\textwidth]{figures_mod/Efficiency_800_900.pdf}
\includegraphics[width=0.49\textwidth]{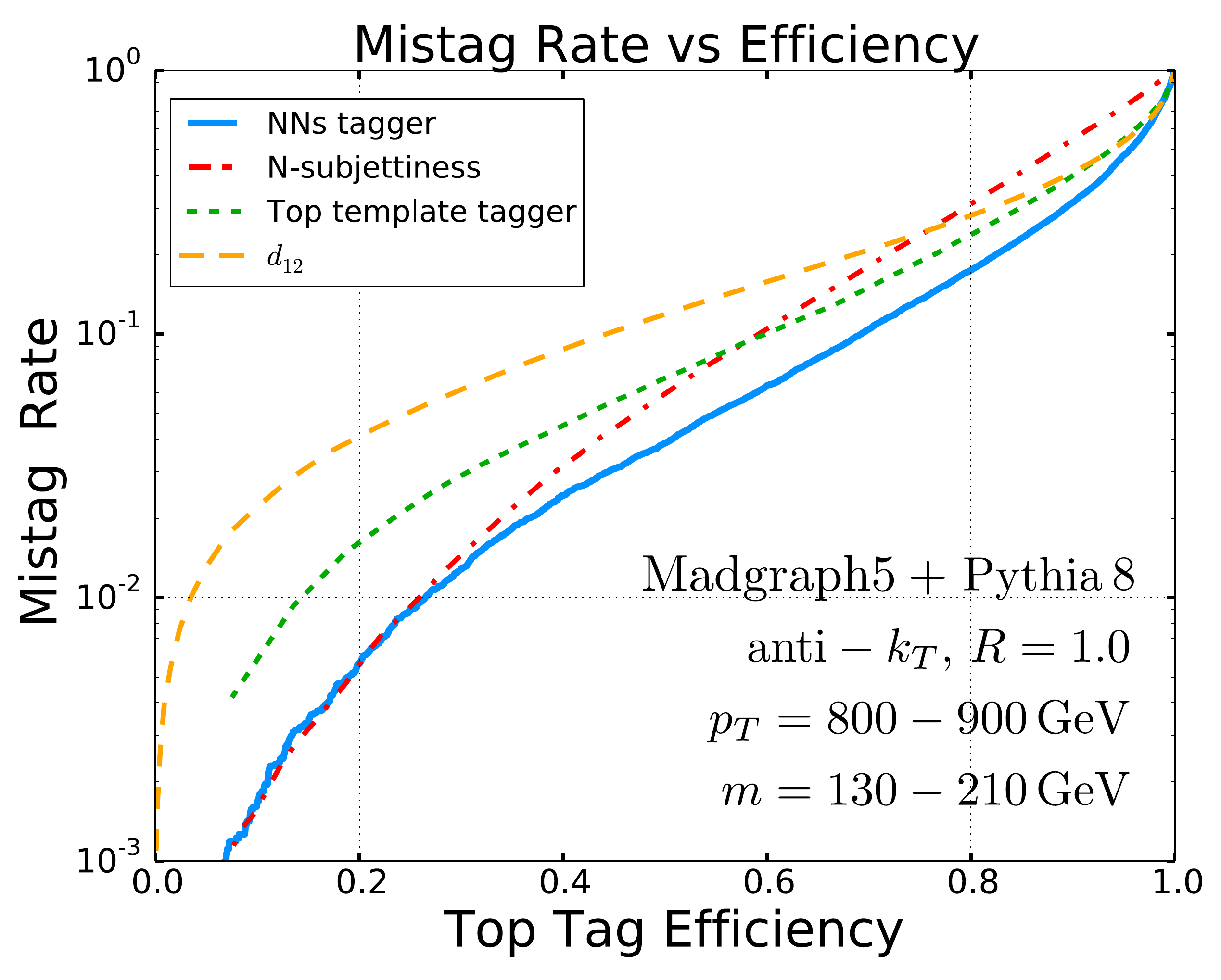}
\end{center}
\caption{Efficiency vs. Mis-tag rate curves for the ANN tagger (blue/solid lines), on jet samples generated with {\tt Pythia 6} (left) and {\tt Pythia 8} (right). For comparison, corresponding curves for three existing top taggers are also shown: 
$d_{12}$ tagger (yellow/dashed), top template tagger (green/dotted), and N-subjettiness (red/dash-dotted). 
\label{fig:pythia8}}
\end{figure}

The final issue we address is the IR-safety of the ANN output. As any observable in jet physics, the ANN score must be IR-safe (or at least Sudakov-safe~\cite{Larkoski:2013paa}) to be useful. Canonically, IR-safety simply requires that the observable be unchanged by exactly collinear $1\to 2$ patron splitting, or an emission of an infinitely soft gluon. Since neither process affects the energy deposits in calorimeter cells $\varepsilon_{ab}$, and since those energy deposits are the only information used by the ANN, its output is manifestly IR-safe by this definition. As a practical matter, however, one might still worry about the sensitivity of the output to non-perturbative physics involved in splittings at small, but finite, angles, and emission of gluons with small, but finite, energy. The modeling of this physics in MC generators such as {\tt Pythia} involves approximations with poorly understood systematic errors, and if the ANN output were determined predominantly by features that depend strongly on the showering model, MC studies would clearly be of very limited utility in assessing the ANN performance on real data. To address this concern, we applied the ANN tagger, trained as described above on jet samples showered with {\tt Pythia 6}, to alternative jet samples generated with the same physics inputs but showered with {\tt Pythia 8}. Showering algorithm of {\tt Pythia 8} differs significantly from {\tt Pythia 6}, in that it incorporates a $p_T$ ordered showering as well as an increased number of underlying event modes and the capability to consider two hard processes. We also applied the three taggers in our comparison pool to the same sample. The result is shown in Fig.~\ref{fig:pythia8}. The ANN tagger continues to perform well on test samples generated with a showering model different from the one used in the training set. This indicates that the features ANN uses to classify jets are physical, rather than artifacts of a particular showering model. Moreover, while there is a non-trivial dependence of the efficiency/mis-tag rate curves on the generator, the effect is of the same size for all taggers considered here. In other words, ANN does not appear to be unusually sensitive in this regard.

\section{Discussion}
\label{sec:discussion}
In this paper, we proposed and explored a new approach to the analysis of jet substructure, specifically top-jet tagging, based on Artificial Neural Network (ANN). The main result of the analysis is captured in Fig.~\ref{fig:EFFvsMT}: the ANN tagger significantly outperforms traditional taggers on the MC ``datasets" used in our study. In a sense, this should not come as a surprise: while the ANN uses the same input information as any other tagger, the training procedure constructs a non-linear function of these inputs which is specifically chosen to maximize its power to classify jets. This maximization takes place on a restricted but extremely broad set of functions, encoded in Fig.~\ref{fig:neural_net} or Eq.~(\ref{NNformula}), and the resulting observable is probably not far away from the theoretical upper limit on classification performance. If this is indeed the case, the ANN can be useful in theoretical studies, serving as a benchmark for other observables used for boosted top tagging. 

Being the first study of this novel approach to top tagging, the analysis presented here does not yet fully capture the complexity of the problem in a realistic experimental environment. The very promising results of this analysis strongly motivate further explorations. Some of the important outstanding issues include:

\begin{itemize}

\item The jets were extracted from event samples including only leading-order SM processes, $t\bar{t}$ and dijet. Subleading processes need to be included. In spite of their smaller rate, they may have outsize effect on the tagger performance: for example, pure QCD processes with high multiplicity of partons in the final state can create ``accidental substructure"~\cite{Hook:2012fd,Cohen:2012yc}, and the ANN would need to learn to distinguish it from real top jets.

\item Pile-up has not been included in our simulations. While many methods to reduce the effects of pile-up have been suggested~\cite{Krohn:2009th,Ellis:2009me}, their interaction with the ANN tagger needs to be explored.

\item Before the method can be applied to real data, concerns about possible MC biases in training the ANN need to be addressed. A preliminary study of this issue suggests that the features that determine the ANN output are not strongly MC-dependent, see Fig.~\ref{fig:pythia8}. However, a more extensive study of this issue is needed, ideally using control/validation samples from real LHC data. In principle, it may even be possible to train the ANN directly on real data, assuming that sufficiently robust training samples can be extracted. This approach would entirely remove concerns about MC biases, and warrants further investigation. 

\end{itemize}  

We plan to address some of these issues in future work.

Another important direction is to further improve the tagger performance. A clear limitation of our tagger is that it only uses HCAL information. Other pieces of information are highly relevant for top tagging, the most obvious one being a sub-jet $b$-tag. This information can certainly be combined with the algorithm presented here to construct an even more powerful tagger. Also, the tagger presented here is based on a rather simple NN architecture and training procedure; more advanced techniques, such as using a convolutional neural network or pre-training the neural network with unsupervised techniques, may result in improved performance. 

Finally, while in this paper we focused exclusively on tops, this approach can equally well be applied to other boosted-object jets, such as $W$ and $h$. It would be interesting to see if performance improvements with respect to traditional taggers can also be achieved in those cases.

In summary, the novel approach to jet tagging based on pattern-recognition techniques, specifically Artificial Neural Networks, shows promise of significant improvements in tagger performance. While the analysis presented in this paper is only the first step, we hope that this approach will eventually become a useful tool in experimental searches for new physics.

\section*{Acknowledgements}
The authors are grateful for conversations with Fabio Maltoni and Jesse Thaler.  We acknowledge the support of the U.S. National Science Foundation through grants PHY-1316222 (MC and MP) and PHY-0844667 (MP). SL is supported in part by the National Research Foundation of Korea grant MEST No. 2012R1A2A2A01045722. MB is supported in part by the Belgian Federal Science Policy Office through the Interuniversity Attraction Pole P7/37. LGA's research leading to these results has received funding from the European Union Seventh Framework Programme (FP7/2007-2013) under grant agreement n$^\circ$ 604102 (HBP).
\begin{appendix}

\section{A Brief Description of Existing Top Taggers}
\label{sec:taggers}


For the purpose of comparison of the ANN tagger to the existing algorithms, we have chosen three existing methods, each one exploiting a different approach to boosted top tagging. In the following list, we give a brief description of the algorithms and the parameters we use for the analysis, while we refer the reader to the references within for detailed discussions.

\begin{itemize}
\item

\textbf{Template Overlap Method (TOM)}: TOM \cite{Almeida:2011aa,Almeida:2010pa, Backovic:2012jj, Backovic:2013bga} is a jet substructure algorithm which aims to match the energy distribution of a fat jet to a partonic structure which models the decay of a heavy boosted particle. TOM algorithm proceeds by comparing libraries of kinematically allowed parton level decays of massive particles (``templates") to the energy distribution of a fat jet. The quality of a match is quantified by the overlap function $Ov$, which minimises the difference between the parton transverse momenta and the amount of $p_T$ deposited in small angular regions around the template patrons (``template sub cones"). An $Ov \sim 1$ score signals a top like jet, while a $Ov \sim 0$ is characteristic of light QCD jets. Here we use the \verb|TemplateTagger v.1.0| \cite{Backovic:2012jk} implementation of the TOM algorithm.

There are many ways generation of template libraries can be implemented. For simplicity and processing speed, here we consider templates at fixed total transverse momentum matched to the mid-point in each fat jet $p_T$ bin of the event samples ($e.g.$ $550$ GeV for fat jet $p_T = 500 - 600$ GeV). We generate the template states using a sequential scan of 40 steps in $\eta, \phi$ over the angular region of $R=1.0$ around the fat jet axis. We match the template libraries to the energy distribution of the fat jet using fixed template sub cones of size $r_3 = 0.1, 0.15, 0.2$ for template $p_T = 1150, 850, 550 \GeV$ respectively, while we allow for the template resolution parameter $\sigma_a = p_{T,\, a} / 3,$ where $p_{T,\,a}$ is the transverse momentum of an individual template parton. 

\item \textbf{N-subjettiness: } Perhaps the most notable example of a ``prong'' tagger is $N$-subjettiness \cite{Thaler:2010tr,Thaler:2011gf}. The algorithm is based on calculating moments $\tau_N$, which serve as estimates of how well the jet energy distribution  can be divided into $N$ regions. The $\tau_N$ are calculated by minimizing the $p_T$ weighted distances between calorimeter energy depositions and trial axes which divide the distribution into $N$ regions, over the space of possible axis configurations. The $N$-subjettiness tagger used in our comparisons is the version publicly available on {\tt HepForge}.\footnote{See {\tt http://fastjet.hepforge.org/contrib/contents/latest.html}}

For the purpose of top tagging the most useful observable is typically the ratio $\tau_3 / \tau_2$, where a high score means that a jet distribution is described better by a three prong configuration. Conversely, a low $\tau_3 / \tau_2$ score is characteristic of two prong jets. Note that in the analysis of this paper we used the angular weight exponent $\beta = 1$ in calculations of $\tau_N$ moments, as suggested in Ref. \cite{Thaler:2011gf}. 

\item
\textbf{ATLAS top tagger}: Jet clustering history can provide useful insight into jet substructure. A notable example is the ATLAS top tagger \cite{Aad:2013nca} which utilises the differences between the top and light jets in the last step of jet clustering. The observable ATLAS uses is $d_{12}$, the value of the  $k_T$ norm at the clustering step which goes from two subjects to one final jet. 
The $d_{12}$ observable is sensitive to the dynamics of hard splittings within the fat jet. The highly asymmetric splittings of typical light jets tend to be characterised by low values of $d_{12}$ with a distribution which falls off sharply with the increase in $d_{12}$, while we expect typical top jets to be characterised by $d_{12} \sim m_t^2/4$.

In addition to $d_{12}$, ATLAS also imposes a lower cut on the trimmed jet mass of $m_j > 130 \GeV$. Unless otherwise noted, here we omit the lower mass cut as the data samples we use for comparison are already restricted to a jet mass window in Eq.~(\ref{eq:mjet}). 

\end{itemize}

\end{appendix}

\bibliography{lit}

\end{document}